\documentstyle[preprint,eqsecnum,aps]{revtex}
\tighten
\begin{document}
\preprint{UCONN-96-10}
\draft
\title{S-matrix elements for gauge theories with and without implemented constraints}
\author{Kurt Haller\thanks{khaller@uconnvm.uconn.edu}} 
\address{Department of Physics, University of Connecticut, Storrs, Connecticut
06269}
\maketitle
\begin{abstract}
We derive an expression for the relation between two scattering transition amplitudes which reflect the 
same dynamics, but which differ in the description of their initial and final state vectors.  In one 
version, the
incident and scattered states are elements of a perturbative Fock space, and solve the eigenvalue problem for 
the `free' part of the Hamiltonian --- the part that remains after the interactions between particle 
excitations have been `switched off'. Alternatively, the incident and scattered states may be 
coherent states that are transforms of these 
Fock states.  In earlier work, we reported on the scattering amplitudes  
for QED, in which a unitary transformation 
relates perturbative and non-perturbative sets of incident and scattered states.  
In this work, we generalize this earlier 
result to the case of transformations that are 
not necessarily unitary and that may not have  unique 
inverses.  We discuss the implication of this relationship for Abelian and non-Abelian gauge theories in 
which the `transformed', non-perturbative states implement constraints, such as Gauss's law.
\end{abstract}
\newpage
\narrowtext

\section{Introduction}
Gauge theories always require the imposition of constraints --- Euler-Lagrange 
equations without time derivatives  (such as Gauss's law) are an example. However, 
when Feynman rules are used, perturbative 
S-matrix elements are evaluated with incident and scattered (`in' and `out') states that do not 
implement these constraints. In particular, in QED, Gauss's law, which couples the 
longitudinal component of the gauge field to charges, is ignored without 
introducing any errors into S-matrix elements.   
More generally, the use of `wrong' 
descriptions of incident and scattered states extends well beyond the 
failure to implement constraints. 
S-matrix elements almost never are evaluated between the exact 
one-particle solutions of the dynamical equations that define even simple models --- most 
gauge theories are far too complicated to permit that.  
We therefore need to understand the circumstances that allow us to dispense with 
the imposition of constraints, as well as with the accurate representation 
of other features of   
incident and scattered states, in evaluating S-matrix elements.
\bigskip
  
In previous work we gave a 
proof that Gauss's law does not need to be implemented in calculating S-matrix elements for 
QED or for other Abelian gauge theories\cite{Abel1,Abel2,Abel3,Abel4}.  The essential idea underlying this proof 
is that Gauss's law for QED is unitarily equivalent to the form that it 
would take for the interaction-free gauge theory, in which the charged sources 
are decoupled from the gauge field. We know, however, that in the case of 
non-Abelian gauge theories, Gauss's law and its limiting form, in which the coupling constant 
that appears in the covariant derivative has been `shut off', 
can not be unitarily equivalent\cite{khymtemp}. In recent work, we have 
constructed states that implement Gauss's law for Yang-Mills theory and QCD
\cite{bellchenhall,chenbellhall}, and have observed that these states can be 
represented as transforms of ordinary perturbative Fock states, but with 
a transformation operator that is not unitary and that cannot even be assumed to be 
non-singular.  It therefore becomes important to extend our earlier proof.  
We now desire a more general result, that relates
S-matrix elements between incident and scattered Fock states to the corresponding  
S-matrix elements for the identical 
dynamical theory, evaluated for states that are transforms of these Fock states. 
But in this case --- in contrast to Refs.~\cite{Abel1,Abel2,Abel3,Abel4} --- 
these transforms will not be assumed to be unitary, nor even to have unique inverses.

\section{Transition amplitudes for perturbative and nonperturbative states.}

We will assume a theory governed by a Hamiltonian 
$H$ that can be represented as 
\begin{equation}
H=H_{0}+H_{\mbox{\scriptsize I}}\,,
\label{eq:pert}
\end{equation}
where $H_0$ describes a `free field' theory, and has a spectrum 
of perturbative eigenstates $|n\,\rangle\,$ which we 
will assume to be elements of a Fock space; $H_{\mbox{\scriptsize I}}$ 
is the part of the 
Hamiltonian that describes interactions among the particle excitations 
which, in general, will include ghost excitation modes. In particular, we will apply 
this formalism to  gauge theories in which $H_0$ describes 
 noninteracting charged particles (we will understand `charge' to include 
electrical charge, color charge, etc.)   and excitations of
gauge fields; and $H_{\mbox{\scriptsize I}}$ describes the interactions 
--- all those parts of 
$H$ that vanish when the coupling constant characteristic of this gauge theory is set $=0.$ 
We will assume that the incident and scattered states of this theory are to be 
represented as a set of nonperturbative eigenstates $|{\bar n}\,\rangle\,,$ of  
the form
\begin{equation}
|{\bar n}\,\rangle={\Xi}\,|n\,\rangle\,.
\label{eq:omega}
\end{equation}
where $\Xi$ is an operator which we will neither assume to be unitary, nor even 
to have a unique inverse. This will enable us to apply the 
work reported in this paper to gauge theories 
in which the state $|n\,\rangle\,$ represents a perturbative state in non-Abelian 
gauge theories 
(for example, a state consisting of 
gluons and quarks) and $|{\bar n}\,\rangle\,$ represents the corresponding state 
that implements the non-Abelian Gauss's law. An explicit construction of 
a transformation operator $\Xi$ for Yang-Mills theory and QCD 
has recently been given\cite{bellchenhall,chenbellhall}.

We will specify two scattering states 
based on these two alternative descriptions of incident particle states  --- the 
perturbative Fock state $|i\,\rangle\,$, 
which obeys $(H_{0}-E_{i}\,)|i\,\rangle=0\,,$  and the nonperturbative state    
$|{\bar i}\,\rangle=\Xi\,|i\,\rangle\,$. These two scattering states are given by\cite{scat}
\begin{equation} 
|\varphi_i\,\rangle = \left[1+ (E_i - H+i\epsilon)^{-1}H_{\mbox{\scriptsize I}}\right]|i\,\rangle 
\label{eq:pertscat}
\end{equation}
and 
\begin{equation} 
|{\bar \varphi}_i\,\rangle = \left[1+ (E_i - H+i\epsilon)^{-1}(H-E_{i}\,)\right]\Xi\,|i\,\rangle   
\label{eq:varnonpert}
\end{equation}
respectively. Our purpose in this work is to find a relation between 
the transition amplitudes 
\begin{equation}
T_{f,i}=\langle\,f\,|H_{\mbox{\scriptsize I}}|{\varphi}_i\,\rangle
,\label{eq:pertamp}
\end{equation}
and 
\begin{equation}
{\bar T}_{f,i}=\langle\,{\bar f}\,|(H-E_{f}\,)
|{\bar \varphi}_i\,\rangle
.\label{eq:nonpertamp}
\end{equation}

We proceed by defining another Hamiltonian, ${\bar H}\,,$ by 
\begin{equation}
\Xi^{\dagger}H\Xi={\bar H}.
\label{eq:Hbar}
\end{equation}
When $\Xi$ is unitary, it is easy to shift $\Xi$ or $\Xi^\dagger$ from one side of Eq.~(\ref{eq:Hbar}) 
to the other. But when $\Xi$ is not unitary, we need to define additional auxiliary quantities to 
achieve that objective.  We define
\begin{equation}
\Xi\,\Xi^\dagger =\alpha=\alpha^{\dagger},
\end{equation} 
and ${\cal A}$ and ${\cal A}^\dagger$ by 
\begin{equation}
{\cal A}\alpha=1\;\;\;\;\;\;\;\mbox{and}\;\;\;\;\;\;\;\alpha{\cal A}^{\dagger}=1.
\label{eq:alpha}
\end{equation} 
We assume that ${\cal A}$ and ${\cal A}^\dagger$ exist, but not that ${\cal A}={\cal A}^\dagger$, 
allowing for the possibility that $\alpha$ may not have a unique inverse.
We also define 
\begin{equation}
{\cal X}=\Xi^{\dagger}{\cal A}^{\dagger}{\cal A}\,\Xi={\cal X}^\dagger.
\end{equation} 
With this expanded set of operators, we easily obtain 
\begin{equation}
\Xi\,{\cal X}=\left[\left(\Xi\;\Xi^{\dagger}\right){\cal A}^{\dagger}\right]{\cal A}\,\Xi\,=
{\cal A}\,\Xi
\label{eq:ksiex}
\end{equation}
and its hermitian adjoint 
\begin{equation}
{\cal X}\Xi^{\dagger}=\Xi^{\dagger}{\cal A}^{\dagger}.
\end{equation}
And, by multiplying Eq.~(\ref{eq:Hbar}) by ${\cal A}\,\Xi$ on the left and using Eq.~(\ref{eq:ksiex}),   
\begin{equation}
H\,\Xi={\cal A}\,\Xi\,{\bar H}=\Xi\,{\cal X}{\bar H}.
\label{eq:Htrans}
\end{equation}
We can transform Eq.~(\ref{eq:varnonpert}) by using Eq.~(\ref{eq:Htrans}) to shift $\Xi$ 
to the extreme left, systematically replacing $H$ by ${\cal X}{\bar H}$ in the wake of the shifted $\Xi\,.$ 
This can be seen clearly from the representation of $(E_i - H+i\epsilon)^{-1}\,\Xi$ as
\begin{eqnarray}
&&\left\{1+\frac{H}{E_i^{(+)}}+\frac{H}{E_i^{(+)}}\cdot\frac{H}{E_i^{(+)}}\cdots+
\left(\frac{H}{E_i^{(+)}}\right)^{n}+\cdots \right\}\left(\frac{\Xi}{E_i^{(+)}}\right)=  
\label{eq:expand} \\
\left(\frac{\Xi}{E_i^{(+)}}\right)&&\left\{1+\frac{{\cal X}{\bar H}}{E_i^{(+)}}+
\frac{{\cal X}{\bar H}}{E_i^{(+)}}\cdot\frac{{\cal X}{\bar H}}{E_i^{(+)}}\cdots+
\left(\frac{{\cal X}{\bar H}}{E_i^{(+)}}\right)^{n}+\cdots \right\}\, \nonumber
\end{eqnarray}
(where $E_i^{(+)}=E_{i}+i\epsilon\,$), which illustrates 
the movement of $\,\Xi\,$ to the left, transforming Eq.~(\ref{eq:varnonpert}) into 
\begin{equation}
|{\bar \varphi}_i\,\rangle =\Xi\left[1+ (E_i - {\cal X}{\bar H}+i\epsilon)^{-1}
({\cal X}{\bar H}-E_{i}\,)\right]|i\,\rangle. 
\label{eq:phitilda}
\end{equation}
Furthermore, in Eq.~(\ref{eq:Htrans}), we can  set ${\bar H}=H_0+{\bar H}_{\mbox{\scriptsize I}}\, $ 
and $\Xi\,{\cal X}{\bar H}_{\mbox{\scriptsize I}}={\bar H}_{\mbox{\scriptsize I}}-
(1-\Xi\,{\cal X}){\bar H}_{\mbox{\scriptsize I}}\,,$ and obtain 
\begin{equation}
{\bar H}_{\mbox{\scriptsize I}}=H_{\mbox{\scriptsize I}}\,\Xi+H_{0}\,{\Xi}-\Xi\,{\cal X}H_{0}+\left(1-\Xi\,{\cal X}\,\right)
{\bar H}_{\mbox{\scriptsize I}}\,.
\label{eq:identa}
\end{equation}
Adding and subtracting $H_0$ from the right hand side of Eq.~(\ref{eq:identa}) leads to 
\begin{equation}
{\bar H}_{\mbox{\scriptsize I}}=H_{\mbox{\scriptsize I}}\,\Xi-H_{0}\left(1-\Xi\right)+
\left(1-\Xi\,{\cal X}\,\right){\bar H};
\label{eq:identb}
\end{equation}
and, after multiplying the hermitian adjoint of Eq.~(\ref{eq:identb}) 
by ${\cal X}$ on the left and 
using Eq.~(\ref{eq:ksiex}) to transform the resulting equation, we obtain
\begin{equation}
{\cal X}{\bar H}_{\mbox{\scriptsize I}}=\Xi^{\dagger}{\cal A}^{\dagger}H_{\mbox{\scriptsize I}}+
\left(1-{\cal X}\right)H_0-
\left(1-\Xi^{\dagger}{\cal A}^{\dagger}\right)H_{0}+{\cal X}{\bar H}\left(1-\Xi^{\dagger}{\cal A}^{\dagger}\right).
\label{eq:identc}
\end{equation}
We can now use Eq.~(\ref{eq:identc}) to obtain
\begin{equation}
\left({\cal X}{\bar H}-E_i\right)|i\,\rangle=\left[\Xi^{\dagger}{\cal A}^{\dagger}H_{\mbox{\scriptsize I}}+
\left({\cal X}{\bar H}-E_i\right)\left(1-\Xi^{\dagger}{\cal A}^{\dagger}\right)\right]|i\,\rangle\,.
\label{eq:Hproject}
\end{equation}
We substitute Eq.~(\ref{eq:Hproject}) into Eq.~(\ref{eq:phitilda});  and,  
in the resulting expression, systematically use ${\cal X}{\bar H}\Xi^{\dagger}=
\Xi^{\dagger}{\cal A}^{\dagger}H\alpha$ to shift $\,\Xi^{\dagger}\,$ to the left  
in the same manner as in Eq.~(\ref{eq:expand}), to obtain 
\begin{equation}
|{\bar \varphi}_i\,\rangle =|i\,\rangle+\alpha\left(E_i - {\cal A}^{\dagger}H\alpha+i\epsilon\right)^{-1}
{\cal A}^{\dagger}H_{\mbox{\scriptsize I}}|i\,\rangle
+i{\epsilon}\,\Xi\left(E_i - {\cal X}{\bar H}+i\epsilon\right)^{-1}\left(1-\Xi^{\dagger}
{\cal A}^{\dagger}\right)|i\,\rangle.
\label{eq:identd}
\end{equation}
Since ${\alpha}{\cal A}^{\dagger}=1$, it is easy to see that 
\begin{equation}
(E_i - {\cal A}^{\dagger}H\alpha+i\epsilon\,)^{-1}
{\cal A}^{\dagger}
={\cal A}^{\dagger}(E_i - H+i\epsilon\,)^{-1};
\end{equation} 
and similarly, we observe that  
\begin{equation}
\Xi\left(E_i - {\cal X}{\bar H}+i\epsilon\,\right)^{-1}\left(1-\Xi^{\dagger}{\cal A}^{\dagger}\right)=
\left(E_i - H+i\epsilon\,\right)^{-1}\left(\Xi-1\right)
\end{equation}
follows from Eq.~(\ref{eq:Htrans}). These two identities can be used to rewrite Eq.~(\ref{eq:identd}) as
\begin{equation}
|{\bar \varphi}_i\,\rangle =\left[1+\left(E_{i}-H+i\epsilon\,\right)^{-1}H_{\mbox{\scriptsize I}}\right]|i\,\rangle
+i\epsilon\,\left(E_{i}-H+i\epsilon\,\right)^{-1}\left(\Xi -1\right)|i\,\rangle.
\label{eq:identf}
\end{equation}
We now use Eq.~(\ref{eq:Htrans}) to rewrite Eq.~(\ref{eq:nonpertamp}) as 
\begin{equation}
{\bar T}_{f,i}=\langle\,f\,|({\bar H}{\cal X}-E_{f}\,)\Xi^{\dagger}
|{\bar \varphi}_i\,\rangle
,\label{eq:Ttilda}
\end{equation} 
as well as to obtain the identity 
\begin{equation}
H_{\mbox{\scriptsize I}}=\Xi\,{\cal X}{\bar H}_{\mbox{\scriptsize I}}-\left(1-\Xi\,{\cal X}\right)H_0 
+H\left(1-\Xi\right)
\label{eq:identalta}
\end{equation} 
and its adjoint
\begin{equation}
H_{\mbox{\scriptsize I}}={\bar H}_{\mbox{\scriptsize I}}{\cal X}\,\Xi^{\dagger}
-H_0\left(1-{\cal X}\,\Xi^{\dagger}\right) 
+\left(1-\Xi^{\dagger}\right)H.
\label{eq:identaltb}
\end{equation}
Finally, we use Eqs.~(\ref{eq:identf}) and (\ref{eq:identaltb}) to transform Eq.~(\ref{eq:Ttilda}) into 
\begin{eqnarray}
{\bar T}_{f,i}&=&T_{f,i}+(E_{i}-E_{f})\langle\,f\,|\left(\Xi^{\dagger}-1\right)\left[|{\varphi}_i\,\rangle+
i\epsilon\,\left(E_{i}-H+i\epsilon\,\right)^{-1}\left(\Xi -1\right)|i\,\rangle\right] 
\label{eq:identaltf} \\
&+&i\epsilon\,\langle\,f\,|\left[\left(\Xi^{\dagger}-1\right)\left(E_{i}-H+i\epsilon\,\right)^{-1}H_{\mbox{\scriptsize I}}+
H_{\mbox{\scriptsize I}}\left(E_{i}-H+i\epsilon\,\right)^{-1}\left(\Xi-1\right)\right]|i\,\rangle  \nonumber \\
&-&i\epsilon\,\langle\,f\,|\left(\Xi^{\dagger}-1\right)\left(\Xi-1\right)|i\,\rangle
+(i\epsilon\,)^{2}\langle\,f\,|\left[\left(\Xi^{\dagger}-1\right)\left(E_{i}-H+i\epsilon\,\right)^{-1}
\left(\Xi-1\right)\right]|i\,\rangle \nonumber
\end{eqnarray}
where $|{\varphi}_i\,\rangle$ is given by Eq.~(\ref{eq:pertscat}).

\section{Discussion.}

When $\Xi$ is unitary, $\alpha=1\,,$ ${\cal A}={\cal A}^{\dagger}=1\,,$ and ${\cal X}=1\,.$ 
The expression in Refs.~\cite{Abel2,Abel3,Abel4}
that relates $T_{f,i}$ and ${\bar T}_{f,i}$ for a unitary transformation, $\Xi\,_{unit}\,,$ 
is given by
\begin{eqnarray}
{\bar T}_{f,i}&=&T_{f,i}+(E_{i}-E_{f})\langle\,f\,|\left(\Xi\,_{unit}^{\dagger}-1\right)|{\varphi}_i\,\rangle  
\label{eq:identunit} \\
&+&i\epsilon\,\langle\,f\,|\left[\left(\Xi\,_{unit}^{\dagger}-1\right)\left(E_{i}-H+i\epsilon\,\right)^{-1}
H_{\mbox{\scriptsize I}}-
{\bar H}_{\mbox{\scriptsize I}}\left(E_{i}-{\bar H}+i\epsilon\,\right)^{-1}\left({\Xi}\,_{unit}^{\dagger}-1\right)\right] 
|i\,\rangle. \nonumber
\end{eqnarray}
When $\Xi$ is the  unitary transformation $\Xi\,_{unit}\,,$ it is straightforward to 
transform Eq.~(\ref{eq:identunit}) so that its form is identical to Eq.~(\ref{eq:identaltf}). The converse does
not hold, however. 
Eq.~(\ref{eq:identunit}) does not describe the relation between $T_{f,i}$ and ${\bar T}_{f,i}$ correctly 
when $\Xi$  is not unitary; auxiliary quantities --- 
$\alpha\,,$ ${\cal A}\,,$ ${\cal A}^{\dagger}\,,$ and ${\cal X}$ --- would be required to avoid quadratic 
$\epsilon$ terms by using barred as well as unbarred Hamiltonians in an equation that resembles 
Eq.~(\ref{eq:identunit}).

It is somewhat surprising that the relation between  $T_{f,i}$ and ${\bar T}_{f,i}$ is so robust, that such a general 
transformation --- one that is neither required to be unitary nor even to have a unique inverse --- makes  
no significant changes in Eq.~(\ref{eq:identaltf}). This robustness may well account for 
our freedom to describe charged particle states quite imperfectly in evaluating perturbative 
S-matrix elements in 
gauge theories. When we use Feynman rules in QED in covariant gauges, we fail to account for the 
electrostatic field that is required by Gauss's law to accompany the 
incident and scattered charged particles.  
Moreover, we similarly omit the `dressing' by transversely polarized propagating 
photons, which account for the magnetic field of a moving charged particle.  Nevertheless, the resulting 
S-matrix elements suffer no harm, save for the infrared divergences 
which stem from the absence of the transversely polarized `soft' photons from the charged 
particle state, and which are 
curable by the Block-Nordsieck algorithm\cite{Bloch}. It is not certain, however, 
that the immunity that applies in QED when we fail to implement Gauss's 
law\cite{Abel1,Abel2,Abel3,Abel4} 
 --- namely, that 
except for the expressions for the renormalization constants,  S-matrix elements are unchanged  when  
 $T_{f,i}$ is substituted for ${\bar T}_{f,i}$  --- also applies to QCD.  

The following remarks 
apply to the relation between ${\bar T}_{f,i}$ and $T_{f,i}$ described by Eq.~(\ref{eq:identaltf}):
The term in Eq.~(\ref{eq:identaltf}) that is proportional to $(E_i-E_f)$ clearly vanishes 
in the S-matrix, which is proportional to $\delta(E_i-E_f)\,.$ The terms in Eq.~(\ref{eq:identaltf}) that are proportional 
to $i\epsilon$ or $(i{\epsilon})^2\,,$ vanish as $\epsilon\rightarrow 0\,,$ unless $(i{\epsilon})^{-1}$ or 
$(i{\epsilon})^{-2}$ appears in a matrix element that has $i\epsilon$ or $(i{\epsilon})^2$ respectively as a coefficient.  
There are various situations in which inverse powers of epsilon can arise in these matrix elements. 
 In one case, the 
matrix element $\langle\,n\,|(\Xi -1)|i\,\rangle$ develops delta-function singularities 
of the form   
\begin{equation}
\langle\,n\,|(\Xi -1)|i\,\rangle=\xi(E_n){\delta}(E_n-E_i)+ \mbox{less singular or non-singular terms}\,,
\end{equation}
 where $\xi(E_n)$ is a 
relatively smooth 
function of $E_n\,.$  An instructive example of $\,i\epsilon\,$ singularities of 
this variety  is the transition amplitude that 
describes nonrelativistic particles scattered by a non-isotropic potential 
represented by $H_{\mbox{\scriptsize I}}\,.$ In this example we will choose $\,\Xi\,$ 
to be the unitary rotation operator $R\,,$ which rotates 
the non-isotropic potential through some set of finite Eulerian angles, 
so that $R$ will commute with $H_0$ but not with $H_{\mbox{\scriptsize I}}\,.$  
In this case ${\bar T}_{f,i}$ is not 
--- and should not be --- identical to $T_{f,i}\,,$ since the former will 
describe a situation in which the non-isotropic potential has been rotated, but the 
incident and scattered states acted on by this potential have remained fixed in space. 
The Green function $\left(E_{i}-H+i\epsilon\,\right)^{-1}$ in 
Eq.~(\ref{eq:identaltf}) can be expanded in the form
\begin{eqnarray}
\left(E_{i}-H+i\epsilon\,\right)^{-1}&=&\left(E_{i}-H_0+i\epsilon\,\right)^{-1}+
\left(E_{i}-H_0+i\epsilon\,\right)^{-1} 
H_{\mbox{\scriptsize I}}\,\left(E_{i}-H+i\epsilon\,\right)^{-1}\\
&=&\left(E_{i}-H_0+i\epsilon\,\right)^{-1}+\left(E_{i}-H+i\epsilon\,\right)^{-1} \nonumber
H_{\mbox{\scriptsize I}}\,\left(E_{i}-H_0+i\epsilon\,\right)^{-1}\,, \nonumber 
\label{eq:green} \nonumber
\end{eqnarray}
and the propagator $\left(E_{i}-H_0+i\epsilon\,\right)^{-1}$ is free to commute with $(R-1)$ and 
$(R^{\dagger}-1)\,$ and to act directly  on $|i\,\rangle$ and $\langle\,f\,|$, 
producing inverse powers of $\,i\epsilon\,.$ We find, in this case, that 
\begin{equation}
\langle\,f\,|\,\left(R^{\dagger}-1\right)\left(E_{i}-H+i\epsilon\,\right)^{-1}
H_{\mbox{\scriptsize I}}|i\,\rangle =(i\epsilon)^{-1}\langle\,f\,|\,\left(R^{\dagger}-1\right)
T\,|i\,\rangle
\label{eq:greenleft}
\end{equation}
where $T=H_{\mbox{\scriptsize I}}+H_{\mbox{\scriptsize I}}
\left(E_{i}-H+i\epsilon\,\right)^{-1}H_{\mbox{\scriptsize I}}\,,$ 
\begin{equation}
\langle\,f\,|H_{\mbox{\scriptsize I}}\left(E_{i}-H+i\epsilon\,\right)^{-1}\,\left(R-1\right)
|i\,\rangle =(i\epsilon)^{-1}\langle\,f\,|\,
T\,\left(R-1\right)|i\,\rangle\,,
\label{eq:greenright}
\end{equation}
and
\begin{equation}
\langle\,f\,|\,\left(R^{\dagger}-1\right)\left(E_{i}-H+i\epsilon\,\right)^{-1}\,\left(R-1\right)
|i\,\rangle =(i\epsilon)^{-2}\left[\langle\,f\,|\,
\left(R^{\dagger}-1\right)T\,\left(R-1\right)|i\,\rangle+{\cal O}(\epsilon )\right]\,,
\label{eq:greenquad}
\end{equation}
so that Eq.~(\ref{eq:identaltf}) for this case, in the limit $E_f{\rightarrow}E_i$ and 
$\epsilon{\rightarrow}0\,,$ does not describe any relationship between ${\bar T}_{f,i}$ and 
$T_{f,i}\,,$ but reduces to the trivial identity 
\begin{equation}
{\bar T}_{f,i}= T_{f,i}+{\bar T}_{f,i}-T_{f,i}\,.
\label{eq:trivial}
\end{equation}  

Eq.~(\ref{eq:identaltf}) can, however, provide useful information about the relation between 
${\bar T}_{f,i}$ and $T_{f,i}$ in QED, when $\Xi\,_{unit}$ is the operator that 
transforms the charged perturbative Fock states into states that implement Gauss's law. 
The unitary operator $\Xi\,_{unit}$ that is used in implementing Gauss's law in QED has 
the form $\Xi\,_{unit}=e^D\,,$ where $D$ is given in earlier work\cite{Abel1,Abel2,Abel3,Abel4}.  
And $(e^D\,-1)$ is a power series in $D\,;$ neither $D$ nor 
$D\,^n\,,$ for any value of $n\,,$ can stand alone in matrix elements between perturbative Fock states that 
describe propagating observable particles --- photons or electrons. Nor does $[\exp{(D)}-1]$ 
commute with $H_0\,;$ $({\bar T}_{f,i}-T_{f,i}\,)$ therefore can not 
have the kind of delta-function singularities, in this case, that 
deprive Eq.~(\ref{eq:identaltf}) of all useful content when a unitary transform $\Xi\,_{unit}$ 
commutes with $H_0\,.$ 
Inverse powers of $i\epsilon$ do arise when 
Eq.~(\ref{eq:identaltf}) is applied to QED, but they arise in a very limited way, and  
 do not invalidate the conclusion that ${\bar T}_{f,i}$ and $T_{f,i}$ are identical, 
although that identity obtains only after changes are made in renormalization 
constants to reflect the fact that $(\Xi\,_{unit}-1)$ and $(\Xi\,_{unit}^{\dagger}-1)$ 
combine with $H_{\mbox{\scriptsize I}}$ to produce additional self-energy insertions 
in external charged particle lines.  These additional self-energy 
insertions change the expressions for renormalization 
constants but do not have any further effect on $S$-matrix elements\cite{Abel1,Abel2,Abel3,Abel4}.

The features of the  unitary transformation operator,  $\Xi\,_{unit}=e^D\,$ that 
is applied to the implementation of Gauss's law in QED limits the 
appearance of inverse powers of $i\epsilon$ 
in matrix elements in Eqs.~(\ref{eq:identaltf}) or~(\ref{eq:identunit}), 
so that only renormalization constants are affected by the substitution of 
$T_{f,i}\,$ for ${\bar T}_{f,i}\,.$ But it is uncertain to what extent this 
immunity from substantial, physically observable differences between $T_{f,i}$ and ${\bar T}_{f,i}$  
extends to the transition amplitude in QCD obtained 
with the non-unitary $\Xi\,$ that implements the non-Abelian Gauss's law. We will not pursue 
this question further here, but will 
leave the detailed application of Eq.~(\ref{eq:identaltf}) 
to the S-matrix in QCD for a later work.

\section{acknowledgments}
The author thanks Mr. M. Belloni for help in checking some algebraic manipulations. 
This research was supported by the Department of Energy
under Grant No. DE-FG02-92ER40716.00.

\end{document}